\documentclass[a4paper,twocolumn,11pt,accepted=2021-10-27]{quantumarticle}
\pdfoutput=1
\usepackage[utf8]{inputenc}
\usepackage[english]{babel}
\usepackage[T1]{fontenc}
\usepackage{amsmath}
\usepackage{hyperref}
\usepackage{bm}
\usepackage[numbers,sort&compress]{natbib}
\usepackage{graphicx}
\usepackage{amssymb,amsmath}
\usepackage{bm}
\usepackage{dcolumn}
\usepackage{subfigure}
\usepackage{float}
\usepackage[OT1]{fontenc} 
\usepackage{slashed}
\usepackage{color}
\usepackage{verbatim}
\usepackage{txfonts}
\usepackage{soul}
\usepackage{subfigure}
\usepackage{comment}
\usepackage{amssymb}

\usepackage{tikz}
\usepackage{lipsum}

\def\bea{\begin{eqnarray}}
\def\eea{\end{eqnarray}}
\def\nn{\nonumber}
\def\ba{\begin{array}}
\def\ea{\end{array}}
\def\nn{\nonumber}
\def\Tr{\text{Tr}}

\def\G{\mathcal{G}}

\begin{document}

	\title{Emergent Replica Conformal Symmetry in Non-Hermitian SYK$_2$ Chains}
	
	\author{Pengfei Zhang}
	\thanks{They contribute equally to this work.}
	\affiliation{Institute for Quantum Information and Matter and Walter Burke Institute for Theoretical Physics, California Institute of Technology, Pasadena, CA 91125, USA}

	\author{Shao-Kai Jian}
	\thanks{They contribute equally to this work.}
	\affiliation{Condensed Matter Theory Center and Joint Quantum Institute,
Department of Physics, University of Maryland, College Park, MD 20742, USA}
	
	\author{Chunxiao Liu}
	\email{chunxiaoliu@ucsb.edu}
	\affiliation{Department of Physics, University of California Santa Barbara, Santa Barbara, CA 93106, USA}
	
	\author{Xiao Chen}
	\email{chenaad@bc.edu}
\affiliation{Department of Physics, Boston College, Chestnut Hill, MA 02467, USA}

\maketitle

\begin{abstract}
		Recently, the steady states of non-unitary free fermion dynamics are found to exhibit novel critical phases with power-law squared correlations and a logarithmic subsystem entanglement. In this work, we theoretically understand the underlying physics by constructing solvable static/Brownian quadratic Sachdev-Ye-Kitaev chains with non-Hermitian dynamics. We find the action of the replicated system generally shows (one or infinite copies of) $\bm{O(2)\times O(2)}$ symmetries, which is broken to $\bm{O(2)}$ by the saddle-point solution. This leads to an emergent conformal field theory of the Goldstone modes. We derive the effective action and obtain the universal critical behaviors of squared correlators. Furthermore, the entanglement entropy of a subsystem $\bm{A}$ with length $\bm{L_A}$ corresponds to the energy of the half-vortex pair $\bm{S\sim \rho_s \log L_A}$, where $\bm{\rho_s}$ is the total stiffness of the Goldstone modes. We also discuss special limits with more than one branch of Goldstone modes and comment on interaction effects. 
\end{abstract}

\section{Introduction}

Under unitary evolution, local quantum information of a generic closed many-body system goes through the process of scrambling and disperses into the entire system. It has long been known that such a process is closely related to thermalization, in which the entanglement entropy of a small subsystem approaches thermal entropy with volume law scaling \cite{Srednicki_1994,Deutsch_1991}. Ever since then, systems that evade quantum thermalization have been of special interest. Several mechanisms have been proposed in an effort to realize such systems, including many-body localization \cite{nandkishore2015many,alet2018many}, prethermalization \cite{abanin2017rigorous,kuwahara2016floquet}, and non-unitary evolution. 

Among these mechanisms, the non-unitary evolution is especially natural since all experimental systems are inevitably open \cite{banerjee2018open}. In a unitary dynamics under repeated measurement, if we follow the {\it quantum trajectories}, the steady state is non-thermal and can exhibit entanglement phase transition from a volume law phase to an area law phase \cite{Li_2018,Cao_Tilloy_2019,Li_2019,Skinner_2019,Chan_2019,Bao_2020,Choi_2020,gullans2019dynamical,gullans2019scalable,jian2019measurementinduced,zabalo2019critical,Tang_Zhu_2020,Szyniszewski_2019,Zhang_2020,goto2020measurementinduced,jian2021yang,buchhold2021effective,bao2021symmetry}. The underlying mechanism also applies to other non-unitary dynamics that host exotic non-thermal phases \cite{sang2020measurement,Lavasani_2021,ippoliti2021fractal,lu2021entanglement,jian2020criticality,Ippoliti_2021,carollo2021emergent}. 
For instance, in free fermion non-unitary systems, it is shown that a stable critical phase exists, in which the entanglement entropy is logarithmic in the subsystem size and the correlation functions in the spatial direction exhibit power-law decay \cite{bao2021symmetry,Chen_2020,alberton2020trajectory,liu2020non}. While these results have been corroborated in numerous numerical simulations, concrete solvable models, in which the entanglement entropy and correlation function can be determined analytically, are still lacking and will prove valuable in the understanding of the intimate relation between quantum thermalization and non-Hermiticity.

The Sachdev-Ye-Kitaev$_q$ (SYK$_q$) models \cite{kitaev2015simple,maldacena2016remarks,Sachdev_Ye} describe $N$ randomly interacting Majorana fermions with infinite-range $q$-fermion interactions. It is found to be solvable in the large-$N$ limit where the entanglement entropy and its quench dynamics can be studied \cite{liu2018quantum,gu2017spread,huang2019eigenstate,10.21468/SciPostPhys.8.6.094,haldar2020Renyi,zhang2020entanglementy,chen2020Replica,liu2020non,shao2021note}. Based on the original SYK model, different generalizations have been constructed. In particular, SYK chains (i.e. coupled SYK dots) have been proposed to understand various dynamical problems in 1D \cite{gu2017local,davison2017thermoelectric,chen2017competition,song2017strongly,zhang2017dispersive,jian2017model,chen2017tunable}. In \cite{saad2018semiclassical,sunderhauf2019quantum}, the authors further consider introducing the temporal randomness into the SYK model, where the static random interaction terms are replaced by Markovian random interactions. This is known as the Brownian SYK model, which is an analogy of the Brownian circuit models \cite{Lashkari_2013,Zhou_Brownian_2019,Xu_2019,Chen_long_range_2019,lucas2019brownian,sunderhauf2019quantum,Piroli_2020}.

In this work, combining ideas of non-Hermiticity and the SYK models, we construct a set of concrete non-Hermitian SYK$_2$ chains. We derive the dynamical properties of the steady states in the large-$N$ limit. By analyzing the saddle-point equation, we explicitly show that for generic values of the coupling parameters, the replicated system exhibits a {\it replica conformal symmetry} due to the existence of Goldstone modes \cite{fn5}. We demonstrate that the Goldstone modes can be probed by the squared correlator, which features the universal critical scaling. We then argue that the entanglement entropy for a subsystem $A$ with length $L_A$ corresponds to the energy of a half-vortex pair \cite{bao2021symmetry}, given by $S\sim \rho_s \log L_A$, where $\rho_s$ is the stiffness of the Goldstone mode \cite{pethick2008bose,zhai_2021}. We finally comment that when the interaction is added, the Goldstone mode acquires a mass, leading to volume-law scaling in the entanglement entropy, which can again be understood from a domain wall picture. 

\section{Model and setup}
	\begin{figure}[tb]
		\centering
		\includegraphics[width=1\linewidth]{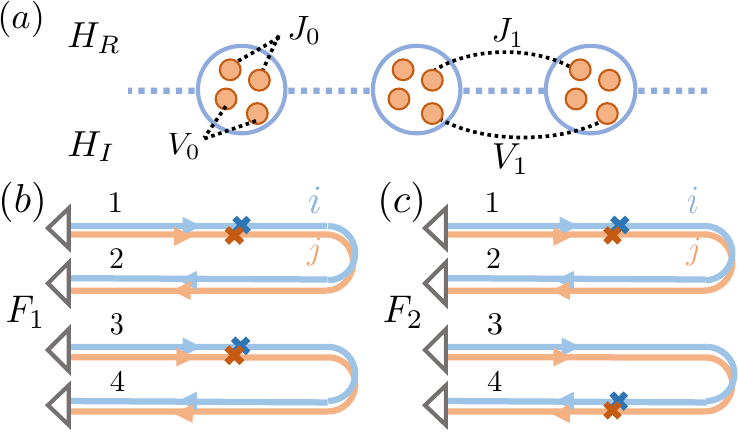}
		\caption{(a). Schematics of the non-Hermitian SYK$_2$ chains studied in the work. The total Hamiltonian is given by $H=H_R-iH_I$ and the interaction can be either static or Brownian, (b/c). The path-integral contour for the squared correlator $F_1/F_2$. Here the triangles represent the contraction with the initial state $|\psi_0\rangle$. The crosses represent the insertion of Majorana operators.}
		\label{fig:schemticas}
	\end{figure}
	
 We consider the non-Hermitian SYK$_2$ chain with a Hamiltonian written in terms of
\begin{equation}
\begin{aligned}
H_R&=\sum_{x,ij}\left[iJ_{ij}^{x,x+1} \chi^i_x\chi^j_{x+1}+i\tilde{J}_{ij}^{x} \chi^i_x\chi^j_x/2\right],\\
H_I&=\sum_{x,ij}\left[iV_{ij}^{x,x+1} \chi^i_x\chi^j_{x+1}+i\tilde{V}_{ij}^{x} \chi^i_x\chi^j_x/2\right].
\end{aligned}
\end{equation}
Here $i=1,2...N$ labels the Majorana modes on each site. The total Hamiltonian for the chain reads $H=H_R-iH_I$. Random hopping parameters $J_{ij}^{x,x+1}$, $\tilde{J}_{ij}^{x}$, $V_{ij}^{x,x+1}$, $\tilde{V}_{ij}^{x}$ are independent Gaussian variables with zero expectation values. We first focus on the case of static correlations and take their variances as 
\begin{equation}
\begin{aligned}
&\overline{(J_{ij}^{x,x+1})^2}=J_1^2/2N,\ \ \ \ \overline{(\tilde{J}_{ij}^{x})^2}=J_0^2/N,\\ &\overline{(V_{ij}^{x,x+1})^2}=V_1^2/2N,\ \ \ \ \overline{(\tilde{V}_{ij}^{x})^2}=V_0^2/N.
\end{aligned}
\end{equation}
 The case of Brownian correlations will be considered in the Appendix \ref{app:Brownian}.

We aim to understand the steady states under the non-Hermitian dynamics. We consider preparing the system in some initial state $|\psi_0\rangle$. For a given disorder configuration, at time $T$, the system is in the state
\begin{equation}
|\psi(T)\rangle=e^{-iHT}|\psi_0\rangle/\sqrt{\left<\psi_0\right|e^{iH^\dagger T}e^{-iHT}\left|\psi_0\right>}.
\end{equation}
We are interested in understanding the emergence of the criticality in the non-unitary dynamics \cite{liu2020non,garcia2021replica}. A natural object to study is the Keldysh equal-time two-point function  $$\left<\chi_x^i\chi_0^j\right>\equiv\frac{\left<\psi_0\right|e^{iH^\dagger T}e^{-iHT/2}\chi_x^i\chi_0^je^{-iHT/2}\left|\psi_0\right>}{\left<\psi_0\right|e^{iH^\dagger T}e^{-iHT}\left|\psi_0\right>}$$ 
evaluated on the steady state with $T\rightarrow \infty$. However, in random systems, the disorder averaged correlator $\overline{\left<\chi_x^i\chi_0^j\right>}$ vanishes, and we should instead consider the disorder averaged squared correlators. We define two types of squared correlators as $$F_1=\sum_{ij}\overline{\left<\chi_x^i\chi_0^j\right>^2}/N,\ \ \ \ \  F_2=\sum_{ij}\overline{\left|\left<\chi_x^i\chi_0^j\right>\right|^2}/N.$$ These correlators measure the fluctuation of two-point functions over disorder realizations. 

The computation of the correlation functions for disordered quantum systems generally requires the introduction of additional disorder replicas. However, for the SYK-like models in the large-$N$ limit, the saddle point solution is diagonal in the disorder replicas \cite{maldacena2016remarks,kitaev2018soft,gu2020notes}. Consequently, $F_1$ and $F_2$ can be represented as four-point functions on the replicated partition function in a single disorder replica:
\begin{equation}
\overline{Z^2}=\overline{(\left<\psi_0\right|e^{iH^\dagger T}e^{-iHT}\left|\psi_0\right>)^2}.
\end{equation} 
In the path-integral formulation of $F_{1,2}$, there are four branches of the evolution contour, as shown in FIG. \ref{fig:schemticas} (b-c). We label these four branches by $1$-$4$ where the evolution is forward on $1,3$ and backward on $2,4$. Following the standard SYK derivation, the system is described by a bilocal $G$-$\Sigma$ action (see Appendix \ref{app:action}), which can then be treated within the saddle-point approximation. The effective action that governs the fluctuations around the saddle point can be subsequently derived, which allows to determine the squared correlators analytically. Moreover, the entanglement calculation can be viewed as introducing defects on the boundary of the contour at time $T$. They excite the fluctuations with an energy increase equal to the entanglement entropy.

\section{Saddle point and symmetry}
We begin with analyzing the saddle-point equation. From now on, we keep the Majorana index $i$ implicitly. For SYK-like models, the saddle-point equation is equivalent to the Schwinger-Dyson equation for $G_x^{ab}(t,t')\equiv\left<\chi^{a}_{x }(t)\chi^{b}_{x}(t')\right>$ with $a,b\in \{1,2,3,4\}$ labeling Majorana fields on different branches of the contour. Viewing $G$ as a matrix in both branch and time space, we have 
\begin{equation}
\begin{aligned}
&\left[(-1)^{a+1}\delta^{ac}\partial_t-\Sigma_x^{ac}\right]\circ G_x^{cb}=I^{ab}.\\
\end{aligned}
\end{equation}
For our static model, the self-energy $\Sigma_x^{ac}$ reads
\begin{equation}
\begin{aligned}
\Sigma_x^{ab}=&(V_1^2-(-1)^{a+b}J_1^2)\frac{G_{x+1}^{ab}+G_{x-1}^{ab}}{2}\\&+(V_0^2-(-1)^{a+b}J_0^2)G_{x}^{ab},
\end{aligned}
\end{equation}
this equation contains large symmetries which are rotations between the $1,3$ or $2,4$ branches. To see this, we define the Fourier transformation as $G_{x}^{ab}(\omega_1,\omega_2)\equiv \int dtdt'~e^{i(\omega_1t+\omega_2t')} G_{x}^{ab}(t,t')$. The Schwinger-Dyson equation is then invariant under $G(\omega_1,\omega_2)\rightarrow O(\omega_1)G(\omega_1,\omega_2)O^{T}(-\omega_2)$, with $O(\omega)=\exp(-\gamma_{13}\theta_{13}^\omega-\gamma_{24}\theta_{24}^\omega)$, here $(\gamma_{cd})^{ab}=\delta_{ac}\delta_{bd}-\delta_{bc}\delta_{ad}$ is a $4\times4$ matrix in the branch space. Consequently, there are infinite copies of the $O(2)\times O(2)$ symmetry, labeled by the frequency $\omega$. Similar symmetry appears when computing the spectral form factor \cite{winer2020exponential}. We note that there is an additional time-reversal symmetry under $t\rightarrow -t$ and $G^{ab}(t,t')\rightarrow(-1)^{a+b}G^{5-a,5-b}(-t,-t')$ \cite{fn1}.

Now we present the solution of the saddle-point equation away from the boundary $t=0$ or $t=T$. Assuming sufficiently large $T$, such that the system reaches the steady state and the Green's function becomes translationally invariant in both space and time directions $G_x^{ab}(t,t')=G_s^{ab}(t-t')$ \cite{fn6}. Since the branches $1,2$ decouple from the branches $3,4$, we expect $G_s^{ab}$ to be block diagonal with $G_s^{13}=G_s^{14}=G_s^{23}=G_s^{24}=0$ and $G_s^{ab}=G_s^{a+2,b+2}$ for $a,b\in \{1,2\}$ \cite{fn2}. In the low-energy limit $|\omega|<\frac{2J^2}{\sqrt{J^2+V^2}}$, $G_s(\omega)$ reads
\begin{equation}\label{eq:GStatic}
\begin{aligned}
G_s^{11}(\omega)&=\frac{i\omega}{2J^2},\\ G_s^{12}(\omega)&=-\frac{1}{2J^2}\sqrt{\frac{4J^4}{J^2+V^2}-\omega^2},
\end{aligned}
\end{equation}
together with $G_s^{22}(t)=-G_s^{11}(t)$ and $G_s^{21}(t)=-G_s^{12}(t)$. Here we have defined $J^2\equiv J_0^2+J_1^2$, $V^2\equiv V_0^2+V_1^2$. For  $|\omega|>\frac{2J^2}{\sqrt{J^2+V^2}}$, we instead have
\begin{equation}
\begin{aligned}
G^{11}_s(\omega)&=i\frac{\omega\left(1-\sqrt{1-\frac{4(J^2-V^2)}{\omega^2}}\right)}{2(J^2-V^2)},\\G^{12}_s(\omega)&=0.
\end{aligned}
\end{equation}

For each frequency $|\omega|<\frac{2J^2}{\sqrt{J^2+V^2}}$, the solutions \eqref{eq:GStatic} break the $O(2)\times O(2)$ symmetry but preserve the time-reversal symmetry. The solutions are only invariant under the symmetry transformation when $\theta_{13}=\theta_{24}$. We denote the residue symmetry group by $O(2)_+$, where the subscript represents the generator is $\gamma_{13}+\gamma_{24}$. The Goldstone mode lives in the coset space $O(2)\times O(2)/O(2)_+$ space and is given by $[G_s,\gamma_{13}-\gamma_{24}]$. In terms of the original bilocal fields, they correspond to $\delta G_{k=0}^{14}(\omega,-\omega)=\delta G_{k=0}^{23}(\omega,-\omega)$. Here we have defined the fluctuation of $G$ fields as $\delta G_{k}^{ab}(\omega_1,\omega_2)\equiv \sum_x~e^{-ikx} \delta G_{x}^{ab}(\omega_1,\omega_2)$. This suggests there should be a line of Goldstone modes labeled by frequency $\omega$. Similar physics has also been observed in \cite{bao2021symmetry}, where only a single $O(2)\times O(2)$ group exists \cite{fn3}.

Special limits exist in which the symmetry of the model is enlarged from $O(2)\times O(2)$. As explained in the Appendix \ref{app:special}, when $J_1=V_0=0$ or $J_0=V_1=0$, rotations between $14$ or $23$ branches are also allowed, which leads to additional Goldstone modes at $\delta G^{13}$ and $\delta G^{14}$. The detailed analysis also shows the mode is near $k=\pi$. However, as we will see later, such modes are not excited when considering a finite subsystem $A$, and thus does not contribute to the entanglement entropy. Nevertheless, as will be explained below, these modes can be probed by the squared correlators.

\section{Effective action and squared correlators}
We now consider fluctuations around the saddle-point solution \eqref{eq:GStatic}. The effective action for fluctuations is given by expanding the $G$-$\Sigma$ action around the low-energy saddle-point solution \eqref{eq:GStatic}. Since we are interested on the low-energy limit, we focus on fluctuations involving two replicas $\{\delta G^{13}$, $\delta G^{14}$, $\delta G^{23}$, $\delta G^{24}\}$. These modes decouple from single-replica modes at the quadratic level. Using the permutation symmetry between two replicas, we combine them into symmetric and anti-symmetric components $\phi_\pm =(\phi_\pm^1,\phi_\pm^2)= \frac1{\sqrt2}(\delta G^{13} \pm \delta G^{24}, \delta G^{14} \pm \delta G^{23}) $. For general coupling parameters, only the symmetric component $\phi_+$ is expected to be gapless due to the existence of Goldstone mode. 

{We are interested in the long-range asymptotic behavior of correlators, which is determined by the effective action at small $k$ and $\omega$. Leaving details of the expansion in Appendix \ref{app:der}, we find that for the static model the effective action of $\phi_+(\omega_1,\omega_2)$ takes the form:}
\begin{equation}
\begin{aligned}
\label{eq:effective_action_static}
&-I_{\text{eff}}/N=\frac{1}{2}\int_{\omega_1\omega_2k}\\&\phi_+
\begin{pmatrix}
2V^2&\frac{i (J^2 + V^2)^{3/2}}{2J^2}\Omega\\
-\frac{i (J^2 + V^2)^{3/2}}{2J^2}\Omega&-\frac{J_1^2+V_1^2}{2}k^2+ \frac{(J^2+V^2)^2\Omega^2}{8J^4}
\end{pmatrix}
\phi_+.
\end{aligned}
\end{equation}
Here we have defined $\Omega=\omega_{1}+\omega_2$. As we expect, there is a line of Goldstone modes at $\Omega=0$ labeled by $\omega=(\omega_1-\omega_2)/2$. Since the solution \eqref{eq:GStatic} is only valid for $|\omega_i|<\frac{2J^2}{\sqrt{J^2+V^2}}$, the effective action \eqref{eq:effective_action_static} is only valid below this cutoff. The theory is a conformal field theory with dynamical exponent $z=1$, as observed in previous numerics for $J_0=V_1=0$ \cite{liu2020non}. It is also useful to reformulate the effective action using the fields $\theta^\omega(\Omega)$ in the coset space. Using $\delta G^{14}=\delta G^{23}=
\sin\theta^\omega(\Omega) G_s^{12}(\omega)\approx \theta^\omega(\Omega) G_s^{12}(\omega)$ and integrating out $\phi_+^1$, we find
\begin{equation}
\label{eq:effective_action_static_theta}
\frac{I_{\text{eff}}}{N}=\frac{1}{2}\int\left(\frac{J_1^2+V_1^2}{J^2+V^2}k^2+\frac{J^2+V^2}{4J^2V^2}\Omega^2\right)|\theta^{\omega}(\Omega,k)|^2.
\end{equation}

The effective action can also be inferred directly from the symmetry perspective: When $\Omega=k=0$, the Goldstone modes $\delta G^{14}$ cost zero energy while $\delta G^{13}$ is gapped. Moreover, under the time-reversal symmetry, we have $\Omega \rightarrow -\Omega$ and $\phi_+\rightarrow \sigma_z \phi_+$, which explains the off-diagonal terms that are linear in frequency. Finally, one can add quadratic terms in $\Omega$ and $k$ allowed by symmetry, which leads to \eqref{eq:effective_action_static}. Consequently, we expect the form of the effective action to be universal. 

This leads to universal scaling for two-point functions of $\phi_+$. In the language of the original fermions, they just correspond to squared correlators $F_1$ and $F_2$. Working out the Fourier transformation, we find the general scaling form
\begin{equation}\label{eq:correlator}
\begin{aligned}
&F_1=-\frac{N\left<\phi_+\phi_+\right>_{11}}{2}\sim\int e^{ikx}\frac{\Omega^2}{\Omega^2+k^2}\sim \frac{1}{x^2}, \\
&F_2=\frac{N\left<\phi_+\phi_+\right>_{22}}{2}\sim\int e^{ikx} \frac{1}{\Omega^2+k^2}\sim \log x.
\end{aligned}
\end{equation}
Here we have omitted parameters that depend on the details of the models. $F_2$ takes the same form as the correlation function of 2$D$ massless bosons, while $F_1$ acquires taking additional derivatives. { The result \eqref{eq:correlator} is under the assumption of disorder replica diagonal saddle-point solutions and the large-$N$ expansion. Comparing to previous works, the scaling of $F_1$ matches the result for the traditional squared correlators evaluated on steady states at small $N$ \cite{Li_2018}. The logrithmic divergence of $F_2$ reflects the fact that the coset field $\theta^\omega(\Omega)$ has a scaling dimension which vanishes in the large-$N$ limit: We consider computing the two-point function of $ G^{12}_s+i\delta G^{14}\sim G^{12}_s e^{i\theta^\omega}$. The result takes the form of $x^{-2\Delta}$, where $\Delta\propto 1/N$ is the scaling dimension of $e^{i\theta^\omega}$. Expanding in terms of $1/N$ gives rise to the logrithmic behavior of $F_2$. }

The result of $F_1$ may be modified if additional Goldstone modes exist. Following a similar route, when there is an additional Goldstone mode at $G^{13}$ and $G^{24}$ (at momentum $\pi$) for special limits, we instead find $F_1\sim (-1)^x\log x$ in the large-$N$ limit.

	\begin{figure}[tb]
		\centering
		\includegraphics[width=1\linewidth]{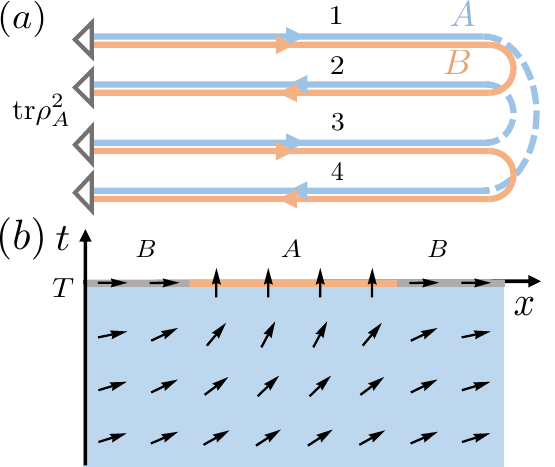}
		\caption{A sketch for the calculation of the second R\'enyi entanglement entropy on the steady state. (a). The path-integral contour for the purity calculation. (b). The effective theory is an XY model on a half-infinite plane, with twisted boundary condition on the boundary $t=T$. The energy increase due to the creation of the half-vortex pair is equal to $S^{(2)}_A(\infty)$.}
		\label{fig:entropy sketch}
	\end{figure}

\section{Entanglement entropy}
We finally turn to the study of the second R\'enyi entanglement entropy of $|\psi(T)\rangle$. We choose the subsystem $A$ as first $L_A$ sites of the chain, and its reduced density matrix $\rho_A(T)=\text{tr}_B\left|\psi(T)\right>\left<\psi(T)\right|$ is obtained by tracing out the remaining part $B$. We consider
\begin{equation}
\begin{aligned}
\text{tr}_A\rho_A^2&=\frac{\text{tr}_A\left(\text{tr}_B~e^{-iHT}\left|\psi_0\right>\left<\psi_0\right|e^{iH^\dagger T}\right)^2}{\left(\left<\psi_0\right|e^{iH^\dagger T}e^{-iHT}\left|\psi_0\right>\right)^2}\\
&\equiv \frac{Z_2(T,A)}{Z^2(T)},
\end{aligned}
\end{equation}
and the second R\'enyi entropy is given by $S^{(2)}_A(T)=-\log\text{tr}_A(\rho_A(T))^2$. Again, by assuming replica diagonality in the disorder replica space \cite{maldacena2016remarks,kitaev2018soft,gu2020notes}, one has
\begin{equation}\label{eq:S}
\overline{S^{(2)}_A(T)}\approx\log \overline{Z_2(T,A)}-\log \overline{Z_2(T,\emptyset)}.
\end{equation}
%Here we have used the fact that $\log \overline{Z_2(T,\emptyset)}\approx2\log \overline{Z}$, which is valid to the leading order of the $1/N$ expansion.

The calculation of entanglement entropy can be mapped to the problem of computing the energy of half-vortex pair that emerges due to the special boundary conditions in the R\'enyi contour \cite{bao2021symmetry}. To illustrate this, we consider the evolution time $T$ to be long enough and focus on the boundary at $t=T$. For subsystem $A$, the twisted boundary conditions connect branches $14$ and $23$ (see FIG. \ref{fig:entropy sketch} (a)) which excites the Goldstone mode and effectively imposes a  $\theta=\pi/2$ vortex on one boundary region of $A$ and a $\theta=-\pi/2$ antivortex on the other. On the contrary, for subsystem $B$, the boundary condition favors $\theta=0$ according to previous analysis. Together with the effective action \eqref{eq:effective_action_static_theta}, the problem maps to (infinite copies of) 2D XY models on a half-infinite plane with additional pinning fields creating half-vortex/antivortex on the boundaries. A sketch for this system is shown in FIG. \ref{fig:entropy sketch} (b). The entanglement entropy \eqref{eq:S} is equal to the energy of this half-vortex-antivortex pair with spatial separation $L_A$, which is known to be $S^{(2)}_A\propto \rho_s \log L_A$ \cite{pethick2008bose,zhai_2021}; such a critical behavior is consistent with previous numerics \cite{liu2020non}. Similarly, the mutual information is equal to (the absolute value of) the interaction energy between two half-vortex pairs, which scales with $1/d^2$ where $d$ is the distance between two subregions. 

\begin{figure}
    \centering
    \includegraphics[width=1\linewidth]{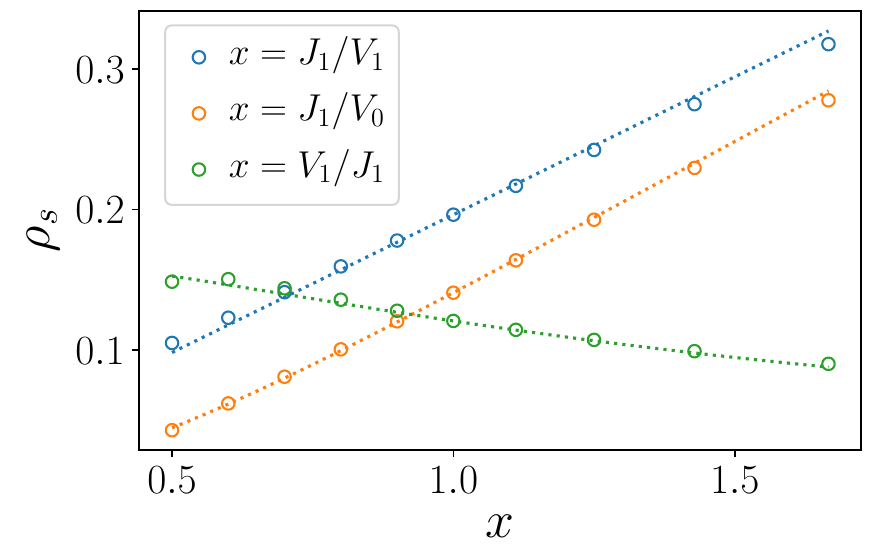}
    \caption{Numerical result for the stiffness $\rho_s$ in the static SYK$_2$ chain. The coupling parameters not mentioned in the plots are set to zero. The analytical expressions \eqref{eq:S_static} is plotted in dashed lines for comparison.}
    \label{fig:stiffness}
\end{figure}

In the large-$N$ limit, the XY model is at
zero-temperature and we can directly apply the mean-field approximation. The stiffness (or the superfluid density) $\rho_s$ is given by summing up contributions from different $\omega$, which gives
\begin{equation}\label{eq:S_static}
S^{(2)}_A\propto \rho_s \log L_A\propto\frac{J}{V}\sqrt{\frac{J_1^2+V_1^2}{J^2+V^2}}N\log L_A.
\end{equation}
Specifically, for large $J_1/V_0$ and $J_0=V_1=0$ the result $J_1/V_0\log L_A$ is consistent with the numerical observation in \cite{liu2020non}. We further perform numerical simulation to verify that the scaling form \eqref{eq:S_static} holds in a large parameter region with finite $J$ and $V$; the numerical result for the stiffness $\rho_s$ is shown in  Fig.~\ref{fig:stiffness} \cite{fn4}.

\section{Connections to circuits at finite \texorpdfstring{$N$}{TEXT}}
The non-unitary random dynamics of free fermion models are also considered in \cite{Chen_2020,alberton2020trajectory,bao2021symmetry} using Gaussian fermionic circuits. In all these works, the steady states exhibit critical phases which have logarithmic entanglement entropy and are interpreted in different approaches. In \cite{Chen_2020}, authors propose a non-linear master equation for squared correlators, which predicts the $1/x^2$ scaling, consistent with the numerics. In \cite{bao2021symmetry}, authors argue the replicated random fermionic circuits can be described by an effective sine-Gordon theory. Different from our result, there is only a single Goldstone mode when the system is in the critical phase. This is due to the absence of energy conservation, which couples fermion modes with different frequencies. Similar behavior can also be observed if we consider the Brownian non-Hermitian SYK$_2$ chains, and we leave this discussion in the Appendix \ref{app:Brownian}.

\section{Discussions}
In this work, we study the non-unitary dynamics of the SYK$_2$ chains for both the static model and the Brownian model. In the replicated space, either system shows (copies) of the $O(2)\times O(2)$ symmetry, which is broken into $O(2)_+$ by the saddle-point solution. The system then obtains an emergent replica conformal symmetry from the fluctuation of Goldstone modes, leading to the universal scaling of squared correlators. The entanglement entropy of a subsystem $A$ can be understood as the energy of a half-vortex pair, which is logarithmic in the subsystem size. 

There are a lot of interesting extensions of this work. Firstly, our derivations can be directly applied to the higher dimensional SYK$_2$ lattices with minor modifications. For 2D lattices, the result is an XY model in 3D. Choosing subsystem $A$ as a circle with perimeter $L_A$, the entanglement entropy then corresponds to the energy of a half-vortex ring, which is proportional to $\rho_SL_A\log L_A$. Secondly, we can add interactions to our model. As an example, consider adding a Brownian SYK$_4$ interaction $\Delta H_R=\sum_{ijkl,x}g_{ijkl}^x(t)\chi^i_x\chi^j_x\chi^k_x\chi^l_x$. This leads to an additional self-energy term $-g\delta(t-t')(-1)^{a+b}G_{x}^{a,b}(t,t')^3$, which breaks the $O(2)\times O(2)$ symmetry explicitly. Consequently, the replicated system becomes gapped, and there is no conformal symmetry anymore. Furthermore, the entanglement entropy corresponds to the energy of a domain wall with length $L_A$, instead of a vortex pair, leading to the volume law. Finally, it is also interesting to construct models with entanglement transitions. Such a model exists if we introduce two copies of the SYK chain and add non-Hermiticity on the fermion parity operator between two copies. The system shows a transition between a critical phase and an area law phase when there is no interaction. After adding interactions, the system shows a second-order transition between a volume-law phase to an area-law phase. These results will be presented in a separate paper \cite{II}.

\textit{Acknowledgment.} We acknowledge helpful discussions with Ehud Altman. PZ acknowledges support from the Walter Burke Institute for Theoretical Physics at Caltech. SKJ is supported by the Simons Foundation via the It From Qubit Collaboration. CL is supported by the NSF CMMT program under Grants No. DMR-1818533. Use was made of computational facilities purchased with funds from the National Science Foundation
(CNS-1725797) and administered by the Center for Scientific Computing (CSC). The CSC is supported by the
California NanoSystems Institute and the Materials Research Science and Engineering Center (MRSEC; NSF
DMR-1720256) at UC Santa Barbara.

\bibliographystyle{unsrtnat}
\bibliography{quantumver.bbl}

\begin{thebibliography}{76}
\bibitem[fn1()]{fn1}
Note that at the level of the Schwinger-Dyson equation, the phase introduced
  here is unnecessary. The choice is to make the symmetry not broken by the
  saddle-point solution.
\bibitem[fn6()]{fn6}
{ We verify this numerically within the disorder replica diagonal assumption by
  solving the Schwinger-Dyson equation at large but finite $T$.}
\bibitem[fn2()]{fn2}
The block diagonal form is verified numerically.
\bibitem[fn3()]{fn3}
This is similar to the Brownian SYK$_2$ chain model.
\end{thebibliography}

\onecolumn\newpage
\appendix

  \section{The path-integral representation of the replicated system}\label{app:action}
  As explained in the main text, we focus on the path-integral representation of the replicated system. The time evolution of a density matrix $\rho$ is
\bea\label{rho}
	\rho(t) = e^{-i H t} \rho e^{i H^\dag t},
\eea
and it does not preserve the normalization. To evaluate $\rho(t)$ using path integral, one needs two contours similar to the Keldysh contour.
Denoted by $1$ and $2$ for forward and backward evolution, the action on these two contours schematically is
\bea
	-I = \int dt \left( \frac12 \chi_x^a (-1)^a \partial_t \chi_x^a + (-1)^a \left( J^{x,x+1}\chi_x^a \chi_{x+1}^a + J^{x} \chi_x^a \chi_{x}^a \right) + i \left( V^{x,x+1} \chi_x^a \chi_{x+1}^a + V^x \chi_x^a \chi_{x}^a \right) \right), \nn \\ 
\eea
where $a = 1, 2$ denotes two contours, and the superscript for Majorana species on each site is suppressed. After introducing two replicas, we aim to compute
  \begin{equation}\label{Z2}
\overline{Z^2}=\overline{(\left<\psi_0\right|e^{iH^\dagger T}e^{-iHT}\left|\psi_0\right>)^2}.
\end{equation}
 This basically doubles the action on the Keldysh contour. Namely, it requires four contours denoted by 1, 2, 3, 4. The effective action on these four contours after integrating out disorder is
\bea \label{eq:full_action}
- \frac{I}N =&& \sum_x \frac12 \Tr \log \Big( (-1)^{a+1} \delta^{ab} \partial_t - \Sigma_x^{ab} \Big)  + \int dtdt'~\Big[ - \frac12 \Sigma_x^{ab} G_x^{ab} \nn \\
&&+ \frac{1}{4} [ V_0^2 (G_x^{ab})^2 + V_1^2 G_x^{ab}G_{x+1}^{ab}]-   \frac{(-1)^{a + b}}4 [J_0^2 (G_x^{ab})^2 +J_1^2 G_x^{ab} G_{x+1}^{ab}] \Big],
\eea
where $G^{ab}$ and $\Sigma^{ab}$ are the bilocal fields with time arguments $t$ and $t'$ omitted, which characterize the two point function of Majarana fermions or the corresponding seld-energy at $a$ and $b$ contours.
As a result, the saddle point equation is 
\bea \label{eq:action}
[G_x^{-1}]^{ab} &=& (-1)^{a+1} \delta^{ab} \partial_t - \Sigma_x^{ab}, \\
\Sigma_x^{ab} &=& [V_0^2 - (-1)^{a+b} J_0^2] G_x^{ab} + [V_1^2 - (-1)^{a+b}J_1^2]  \frac{ G_{x-1}^{ab} + G_{x+1}^{ab} }2.
\eea

  \section{The derivation of the effective action}\label{app:der}
The effective action is given by expanding the $G-\Sigma$ action \eqref{eq:full_action} or \eqref{eq:BSYK2_action} around the saddle-point solution. In this section, we give a detailed derivation for the effective actions. 

\subsection{The effective action of the static model}
We consider the saddle point fluctuations,
\bea
	\Sigma_x(\omega_1, \omega_2) = \Sigma_s(\omega_1) 2\pi \delta(\omega_1 + \omega_2) + \delta \Sigma_x(\omega_1, \omega_2).
\eea
and similar for $G$, where we have used the convention $\Sigma(t_1, t_2) = \int \frac{d\omega_1}{2\pi} \frac{d\omega_2}{2\pi} \Sigma(\omega_1, \omega_2) e^{-i \omega_1 t_1 - i \omega_2 t_2}$ for Fourier transformation.  
The trace log term in (\ref{eq:full_action}) gives rise to [as the first line in (\ref{eq:full_action}) does not depend on lattice site, we suppress lattice index $x$ until we move on to evaluate the second line in (\ref{eq:full_action})]
\bea
- \frac14 \int_{\omega_1, \omega_2} \delta \Sigma^{ a b}(\omega_1, \omega_2) G_s^{d a}(\omega_1) G_s^{ b  c}(-\omega_2) \delta \Sigma^{ cd}(-\omega_2, -\omega_1).
\eea
where we have defined $\int_\omega \equiv \int \frac{d\omega}{2\pi}$ for conciseness.
We will be interested in the correlation between $1$, $2$ and $3$, $4$, so we focus on these correlation functions.
Assuming the symmetry $\Sigma^{ a b} (t_1, t_2) = -\Sigma^{ b a}(t_2, t_1)$, we can bring the kernel into
\bea
	 \int_{\omega_1, \omega_2} \frac{1}{8J^4} \hat \sigma(\omega_1, \omega_2) 
	 \left( \ba{cccc} \omega_1 \omega_2 & - f(\omega_1) f(\omega_2) & -i\omega_1 f(\omega_2) & -i \omega_2 f(\omega_1) \\ 
	  - f(\omega_1) f(\omega_2) & \omega_1 \omega_2 &  -i\omega_2 f(\omega_1) & -i \omega_1 f(\omega_2) \\
	  i\omega_1 f(\omega_2) & i \omega_2 f(\omega_1) & - \omega_1 \omega_2 & f(\omega_1) f(\omega_2) \\
	  i\omega_2 f(\omega_1) & i \omega_1 f(\omega_2) & f(\omega_1) f(\omega_2) & -\omega_1 \omega_2 \ea\right) 
	  \hat \sigma(-\omega_1, - \omega_2), \nn \\
\eea
where we have defined $\hat \sigma \equiv ( \delta \Sigma^{13}, \delta\Sigma^{24}, \delta\Sigma^{14}, \delta\Sigma^{23} )$
and $f(\omega) \equiv \sqrt{\frac{4J^4}{J^2 + V^2} - \omega^2}$.

If one notices the coupling term $ - \frac12 \int \Sigma^{ a b} G^{ a b} $ in (\ref{eq:full_action}), then it is a straightforward task to integrate out $\hat \sigma$ field, and the resultant action reads
\bea \label{eq:frequency}
	&& \int_{\omega_1, \omega_2} \frac{(J^2 + V^2)^2}{8J^4}  \nn \\
	&& \times \hat g_k(\omega_1, \omega_2) 
	\left( \ba{cccc} -\omega_1 \omega_2 &  f(\omega_1) f(\omega_2) & i\omega_1 f(\omega_2) & i \omega_2 f(\omega_1) \\ 
	f(\omega_1) f(\omega_2) & -\omega_1 \omega_2 &  i\omega_2 f(\omega_1) & i \omega_1 f(\omega_2) \\
	-i\omega_1 f(\omega_2) & -i \omega_2 f(\omega_1) &  \omega_1 \omega_2 & -f(\omega_1) f(\omega_2) \\
	-i\omega_2 f(\omega_1) & -i \omega_1 f(\omega_2) & -f(\omega_1) f(\omega_2) & \omega_1 \omega_2 \ea\right) 
	\hat g_{-k}(-\omega_1, -\omega_2),
\eea
where $\hat g_k \equiv ( \delta G^{13}_k, \delta G^{24}_k, \delta G^{14}_k, \delta G^{23}_k )$.
We restore the lattice index by going to the momentum space using $g_k \equiv \frac1{\sqrt{L}} \sum_x g_x e^{-i k x} $ with $L$ the number of sites.

The second line in (\ref{eq:full_action}) is simple, and leads to 
\bea \label{eq:momentum}
	&& \int_{\omega_1, \omega_2} \frac12 \hat g_k (\omega_1, \omega_2) 
	\left( \ba{cccc} V_k^2 - J_k^2 & 0 & 0 & 0 \\ 
	0 & V_k^2 - J_k^2 &  0 & 0 \\
	0 & 0 &  V_k^2 + J_k^2 & 0 \\
	0 & 0 & 0 & V_k^2 + J_k^2 \ea\right) 
	\hat g_{-k}(-\omega_1, -\omega_2).
\eea
So the effective action is given by the sum of (\ref{eq:frequency}) and (\ref{eq:momentum}).

The fields $\delta G^{14}$ and $\delta G^{23}$ are related to a Goldstone mode. 
More precisely, $\delta G^{14} + \delta G^{23}$ is the Goldstone mode, so they are gapless at zero frequency and momentum.
Expanding the effetive action near zero frequency and keeping the leading term, it reads
\bea \label{eq:effective_action_staticapp}
	&& \frac{-I_{\text{eff}}}{N} = \notag\\
	&&\frac12\int_{\omega_1, \omega_2}  \hat \phi_+  
    \left( \ba{cccc} J^2 - J_k^2 + V^2 + V_k^2 - \frac{(J^2+V^2)^2(\omega_1 + \omega_2)^2}{8J^4} & \frac{i (J^2 + V^2)^{3/2}}{2J^2} (\omega_1 + \omega_2 ) \\
	\frac{-i (J^2 + V^2)^{3/2}}{2J^2} (\omega_1 + \omega_2 ) & -J^2 + J_k^2 - V^2 + V_k^2 + \frac{(J^2+V^2)^2(\omega_1 + \omega_2 )^2}{8J^4} \ea \right) \hat \phi_+
	 \nn \\ 
	 && + \hat \phi_-
	  \left( \ba{cccc}
	 - J^2 - J_k^2 - V^2 + V_k^2 + \frac{(J^2+V^2)^2(\omega_1 - \omega_2 )^2}{8J^4} & \frac{i (J^2 + V^2)^{3/2}}{2J^2} (\omega_1 - \omega_2 ) \\
	 \frac{-i (J^2 + V^2)^{3/2}}{2J^2} (\omega_1 - \omega_2 ) &  J^2 + J_k^2 + V^2 + V_k^2 - \frac{(J^2+V^2)^2(\omega_1 - \omega_2 )^2}{8J^4} \ea \right) \hat \phi_-,
\eea 
where we make the symmetrization and antisymmetrization of the fields $\hat \phi_\pm = \frac1{\sqrt2}(\delta G^{13} \pm \delta G^{24}, \delta G^{14} \pm \delta G^{23}) $, like the conventional Kelydsh rotation. 
And remember  $J_k^2 = J_0^2 + J_1^2 \cos k$, $V_k^2 = V_0^2 + V_1^2 \cos k$. Only keeping the $\hat \phi_+$ part and expand each element to the leading order in small $k$ and $\omega$, we get the effective action presented in the main text.

\subsection{Relation to the entanglement entropy}
Here we explain the boundary condition for $\theta$ when computing the entanglement entropy. As discussed in the main text, the entanglement entropy corresponds to
\begin{equation}
e^{-S^{(2)}_A}=\text{lim}_{T\rightarrow \infty}\text{tr}_A\left(\text{tr}_B(e^{-iHT}|\psi_0\rangle \langle \psi_0|e^{iH^\dagger T})\text{tr}_B(e^{-iHT}|\psi_0\rangle \langle \psi_0|e^{iH^\dagger T})\right).
\end{equation}
In other words, for sites in the subsystem $B$, the boundary condition connects the forward and the backward evolution in the same replica which favors $G^{12}\neq 0$ and $G^{14}= 0$, while for sites in the subsystem $A$, the boundary condition connects the forward and the backward evolution in different replicas which favors $G^{12}= 0$ and $G^{14}\neq 0$. This directly corresponds to the excitation of Goldstone modes. More explicitly, if we work out the low-energy manifold 
\begin{equation}
G_x(t)=
\begin{pmatrix}
G^{11}_s&\cos (\theta)G^{12}_s&0&\sin(\theta)G^{12}_s\\
-\cos (\theta) G^{12}_s&-G^{11}_s&\sin(\theta)G^{12}_s&0\\
0&-\sin(\theta)G^{12}_s&G^{11}_s&\cos (\theta)G^{12}_s\\
-\sin(\theta)G^{12}_s&0&-\cos (\theta)G^{12}_s&-G^{11}_s
\end{pmatrix}.
\end{equation}
Then in $B/A$ subsystem, the boundary condition favors $\theta=0$ or $\pi/2$. Computing the entropy corresponds to the calculation of the energy of two half-vortexes with opposite charge.

  \section{The Brownian non-Hermitian SYK\texorpdfstring{$_2$}{TEXT} chains}\label{app:Brownian}
We can also generalize our SYK$_2$ chain to the Brownian case. Now, we treat $J_{ij}^{x,x+1}$, $\tilde{J}_{ij}^{x}$, $V_{ij}^{x,x+1}$, $\tilde{V}_{ij}^{x}$ as time-dependent variables with 
\begin{equation}
\begin{aligned}
&\overline{J_{ij}^{x,x+1}(t)J_{ij}^{x,x+1}(0)}=\frac{J_1}{2N}\delta(t),\ \ \ \ \ \ \ \ \ \ \ \ \overline{\tilde{J}_{ij}^{x}(t)\tilde{J}_{ij}^{x}(0)}=\frac{J_0}{N}\delta(t),\\&\overline{V_{ij}^{x,x+1}(t)V_{ij}^{x,x+1}(0)}=\frac{V_1}{2N}\delta(t),\ \ \ \ \ \ \ \ \ \ \ \overline{\tilde{V}_{ij}^{x}(t)\tilde{V}_{ij}^{x}(0)}=\frac{V_0}{N}\delta(t).
\end{aligned}
\end{equation}

Since the Hamiltonian is time-dependent, and we keep the time-ordering in \eqref{rho} and \eqref{Z2} implicit. Following similar steps, the result reads
\bea \label{eq:BSYK2_action}
- \frac{I}N &=& \sum_x\frac12 \Tr \log \left( (-1)^{ a+1} \delta^{ a b} \partial_t - \Sigma_x^{ a b} \right) + \int dtdt'~ \Big[ - \frac12 \Sigma_x^{ a b} G_x^{ a b} \nn\\
	&& 	+ \frac{ \delta}{4} [V_0 (G_x^{ a b})^2 + V_1 G_x^{ a b} G_{x+1}^{ a b}]- \frac{(-1)^{ a +  b}\delta}{4}  [ J_0 (G_x^{ a b})^2 + J_1 G_x^{ a b} G_{x+1}^{ a b} ] \Big], 
\eea
Here we have ommited the time arguments for bilocal fields as in the static case. The main difference compared to the static case is the appearance of the additional $\delta=\delta(t_1-t_2)$ due to the lack of correlation in the time direction. This gives 
\bea
	[G_x^{-1}]^{ a b} &=& (-1)^{ a+1} \delta^{ a b} \partial_t - \Sigma_x^{ a b}, \\
\Sigma_x^{ a b} &=& \delta[V_0 - (-1)^{ a+  b} J_0]  G_x^{ a b} + \delta[V_1 - (-1)^{ a+ b} J_1] \frac{ G_{x-1}^{ a b} + G_{x+1}^{ a b}}{2}.
\eea

Different from the static case, now there is only a single $O(2)\times O(2)$ symmetry, given by the transformation $G(t,t')\rightarrow OG(t,t')O^{T}$ with $O=\exp(- c_{13}\theta_{13}- c_{24}\theta_{24})$. The saddle-point solution takes a simpler form
\begin{equation}\label{eq:GBrownian}
G_s^{11}(\omega)=\frac{i\omega}{\omega^2+\Gamma^2/4},\ \ \ \ G_s^{12}(\omega)=-\frac{\Gamma/2}{\omega^2+\Gamma^2/4}.
\end{equation}
Here $\Gamma=V+J= V_0+V_1+J_0+J_1$ is the quasi-particle decay rate. Similar to the static case, the saddle-point solution also breaks the symmetry down to $O(2)_+$, leading to a single Goldstone mode.

Now we consider saddle-point fluctuations,
\bea
	\Sigma(t_1, t_2) = \Sigma_s(t_1, t_2) + \delta \Sigma(t_1) \delta(t_{12}), \quad G(t_1, t_2) = G_s(t_1, t_2) + \delta G(t_1, t_2).
\eea
We first evaluate the $\Tr \log$ term and then the interaction terms. 
Expanding the $\Tr \log$ term into second order, we arrive at
\bea
	&& - \frac14 \int_{\omega, \Omega} \Tr\left[ G_s(\omega + \Omega) \delta \Sigma(\Omega) G_s(\omega) \delta \Sigma(-\Omega) \right]
	= \frac12 \int_{\Omega} \sigma^{T}(\Omega) \mathcal{M} \sigma(-\Omega),
\eea
where $\int_\Omega \equiv \int \frac{d\Omega}{2\pi} $, and we have used the Fourier transform $\delta \Sigma(\Omega) = \int dt \delta \Sigma(t) e^{i \Omega t}$. 
In evaluating the kernel, we have used the symmetry of bilocal field $\Sigma^{ a b}(t_1, t_2) = - \Sigma^{ b a}(t_2, t_1)$, so there are four independent diagonal fields and six independent off-diagonal fields.
The full kernel implies that (a) off-diagonal fields decouple from diagonal field and (b) four out of six independent off-diagonal fields have nontrivial interactions.
We denote these four nontrivial off-diagonal fields to be $\sigma = \left( \delta \Sigma^{13}, \delta \Sigma^{24} ,\delta \Sigma^{14},\delta \Sigma^{32} \right)^T$, and the corresponding kernel reads
\bea
	\mathcal{M} = \frac{\Gamma}{2(\Omega^2 + \Gamma^2)} \left( \ba{cccc} -1 & -1 & -i & i \\
	 -1 & -1 & -i & i \\
	  i & i & 1 & -1 \\
	  -i & -i & -1 & 1 \ea \right). %\quad a = .
\eea
It is easy to check that the kernel has two zero modes, and we can make redefinition of the fields
\bea
	\varphi(-\Omega) = U \sigma(-\Omega ) , \quad  U = \frac1{\sqrt2}  \left( \ba{cccc} 1 & 1 & 0 & 0 \\
	0 & 0 & -1 & 1 \\
	-1 & 1 & 0 & 0 \\
	0 & 0 & 1 & 1 \ea \right),
\eea
such that the last two fields are zero modes and the quadratic action for the first two fields are (we use $\varphi$ to denote the first two fields in the following)
\bea
	\frac12 \int_{\Omega} \varphi^{T}(\Omega) \mathcal{M}_U \varphi(-\Omega), \quad 
	\mathcal M_U = U \mathcal M U^\dag = \left( \ba{cccc} - \frac{\Gamma}{\Omega^2 + \Gamma^2} & \frac{i \Omega}{\Omega^2 + \Gamma^2} \\
	-\frac{i \Omega}{\Omega^2 + \Gamma^2} & -\frac{\Gamma}{\Omega^2 + \Gamma^2} \ea \right).
\eea

\begin{figure}
    \centering
    \includegraphics[width=0.5\linewidth]{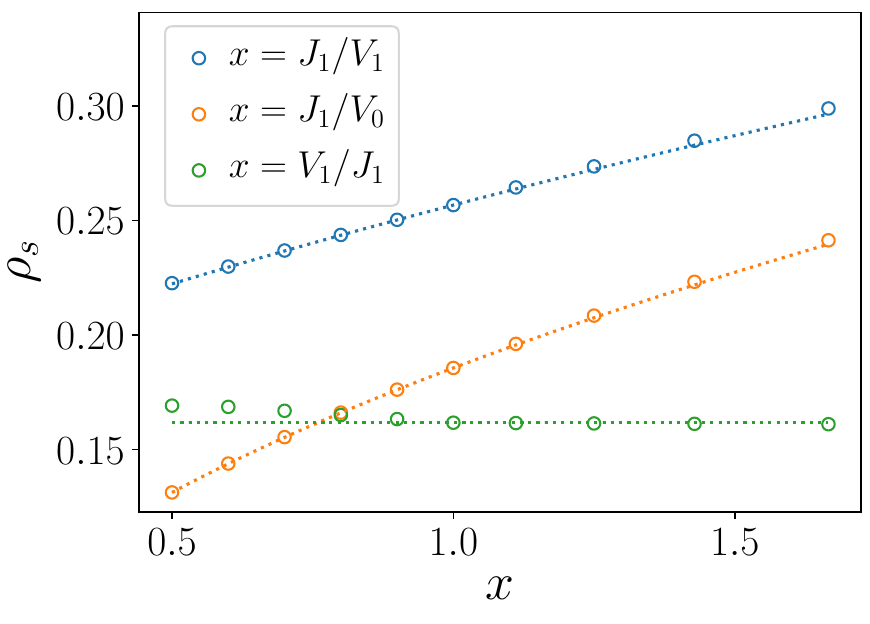}
    \caption{Numerical result for the stiffness $\rho_s$ in the Brownian SYK$_2$ chain. The coupling parameters not mentioned in the plots are set to zero. The analytical expressions \eqref{eq:S_Brownian} is plotted in dashed lines for comparison.}
    \label{fig:stiffness2}
\end{figure}

Now we are ready to integrate out $\varphi$ fields. 
For the two zero modes, because they couple linearly to $\delta G$ fields in the action~(\ref{eq:BSYK2_action}), integrating them out leads to two constraints.
\bea
	\delta G^{13}(t,t) = \delta G^{24}(t,t), \ \ \ \ \ \  \delta G^{14}(t,t) =- \delta G^{32}(t,t).
\eea
For notational simplicity, we introduce another fields 
\bea
	\phi_1(\Omega) =\sqrt{2}\int dt ~\delta G^{13}(t,t) e^{i \Omega t}, 
	\ \ \ \ \ \ \phi_2(\Omega) = -\sqrt{2}\int dt ~\delta G^{14}(t,t) e^{i \Omega t}.
\eea
In terms of these field of interest $\varphi_i, \phi_i$, $i=1,2$, the quadratic action reads
\bea
	-\frac{I_{\text{eff}}}N =	\frac{1}{2}\sum_x \int_{\Omega} \left( \frac12 \varphi_x^{T}(\Omega) \mathcal{M}_U \varphi_x(-\Omega) - \frac{\sqrt2}2 \left[ \phi_x(\Omega)\varphi_x(-\Omega) + \varphi_x(\Omega)\phi_x(-\Omega)  \right] \right)\nn \\
	+ \frac{1}{2}\int_{\Omega k} \left( V_k(\phi_{1,k} \phi_{1,-k} + \phi_{2,k}\phi_{2,-k})  - J_k (\phi_{1,k} \phi_{1,-k}- \phi_{2,k}\phi_{2,-k}) \right).
\eea
where we restore the space index and make Fourier transform to momentum space in the second line, $ \phi_x = \frac1{\sqrt{L}} \sum_k \phi_k e^{i k x}$, $L$ is the number of the site of the chain. 
And we also define $J_k = J_0 + J_1 \cos k$, $V_k = V_0 + V_1 \cos k$, and one should distinguish it from the case of regular SYK$_2$ model. It is then straightforward to integrate out $\varphi$ field to have
\bea \label{eq:effective}
		-\frac{I_{\text{eff}}}N = \frac{1}{2} \int_{\Omega k} \phi_k(\Omega) \left( \ba{cccc} J-J_k + V + V_k & -i \Omega \\ i\Omega & -J + J_k - V + V_k\ea \right) \phi_{-k}(-\Omega).
\eea
Expanding each element to the leading order in small $k$ and $\omega$, we get 
\begin{equation}
\begin{aligned}
\label{eq:effective_action_Brownian}
&-\frac{I_{\text{eff}}}{N}=\frac{1}{2}\int_{\Omega k}\phi_k(\Omega)
\begin{pmatrix}
2V&i\Omega\\
-i\Omega&-(J_1+V_1)k^2/2
\end{pmatrix}
\phi_k(\Omega).
\end{aligned}
\end{equation}
In terms of the coset space variable $\theta(\Omega)$, this becomes the effective action of a single Goldstone mode
\begin{equation}
\label{eq:effective_action_Brownian_theta}
\frac{I_{\text{eff}}}{N}=\frac{1}{2}\int_{\Omega k}\left(\frac{J_1+V_1}{4}k^2+\frac{1}{4V}\Omega^2\right)|\theta(\Omega,k)|^2,
\end{equation}
which takes a similar form as the static case due to the time-reversal symmetry. Consequently, the universal scaling of the squared correlators still applies and $\rho_s$ can be extracted from \eqref{eq:effective_action_Brownian_theta} to obtain
\begin{equation}\label{eq:S_Brownian}
\text{Brownian:}\ \ \ \ \ S^{(2)}_A\propto \rho_s \log L_A=\sqrt{\frac{J_1+V_1}{V}}N\log L_A.
\end{equation}
Such a scaling is again verified numerically; the numerical value of the stiffness $\rho_s$ is shown in Fig. \ref{fig:stiffness2}.

  \section{Discussions on special limits of the model}\label{app:special}
As mentioned in the main text, in special limits, the symmetry of the system can be larger and additional Goldstone modes appear. In this section, we give a detailed discussions on these cases. We take the static case as an example, and finally comment on the Brownian case. 

For the saddle-point equation \eqref{eq:action}, we consider:
\begin{enumerate}
\item In special cases the symmetry of the equation can be larger than $O(2)\oplus O(2)$ (for each frequency $\omega$). For example, when we have purely Hermitian evolution with $V_0=V_1=0$, we can define $\tilde{G}^{ab}=(-1)^a\tilde{G}^{ab}$. The self-consistent equation then reads 
\begin{equation}
\begin{aligned}
&i\omega_1\tilde{G}_x^{ac}(\omega_1,-\omega_2)-\int \frac{d\omega_3}{2\pi}\tilde{\Sigma}_x^{ab}(\omega_1,-\omega_3) \tilde{G}_x^{bc}(\omega_3,-\omega_2)=I^{ac}\delta(\omega_1-\omega_2),\\
&\tilde{\Sigma}_x^{ab}(\omega_1,\omega_3)=-J_1^2\frac{\tilde G_{x+1}^{ab}(\omega_1,\omega_3)+\tilde G_{x-1}^{ab}(\omega_1,\omega_3)}{2}-J_0^2\tilde G_{x}^{ab}(\omega_1,\omega_3).
\end{aligned}
\end{equation}
The equation now becomes $O(4)$ symmetric: $\tilde{G}_x^{ac}(\omega_1,\omega_2) \rightarrow O(\omega_1)\tilde{G}_x^{ac}(\omega_1,\omega_2)O^T(-\omega_2)$ with arbitrary $O(4)$ matrices $O(\omega)$. The effective action now leads to diffusive behavior 
\begin{equation}
\left<\delta G^{13}\delta G^{13}\right>=-\left<\delta G^{24}\delta G^{24}\right>\sim \frac{1}{\Omega^2+k^4}.
\end{equation}
where we have dropped non-universal coefficients.

\item There is additional symmetry between the forward and the backward evolution if $J_0=0$ and $V_1=0$. The reason is that if we consider two evolutions and neglect their boundary conditions
\begin{equation}
e^{-iHT}e^{iH^\dagger T}=e^{-iH_RT-H_IT}e^{iH_RT-H_IT}
\end{equation}
If we define $\chi_x'=\chi_x(-1)^x$, we find $H_R\rightarrow -H_R$ with $H_I$ unchanged. Consequently, There is an additional Goldstone mode with momentum $\pi$. The mode appears at $G^{13}_{k=\pi}(\omega,\omega)$ and $G^{24}_{k=\pi}(\omega,\omega)$. Explicitly, we have 
\begin{equation}
\left<\delta G^{13}\delta G^{13}\right>= \frac{1}{\omega^2+(k-\pi)^2}.
\end{equation}

\item Similar modes exists when we instead have $V_0=0$ and $J_1=0$. The idea is again to perform the transformation $\tilde{G}^{a,c}=(-1)^a\tilde{G}^{a,c}$. The self-energy becomes:
\begin{equation}
\tilde{\Sigma}_x^{ab}(\omega_1,\omega_3)=V_1^2(-1)^{a+b}\frac{\tilde G_{x+1}^{ab}(\omega_1,\omega_3)+\tilde G_{x-1}^{ab}(\omega_1,\omega_3)}{2}-J_0^2\tilde G_{x}^{ab}(\omega_1,\omega_3)
\end{equation}
Back to the evolution operator language, this corresponds to the self-consistent equation of 
\begin{equation}
e^{-iH_RT-H_IT}e^{-iH_RT+H_IT}e^{-iH_RT-H_IT}e^{-iH_RT+H_IT}
\end{equation}
Here we neglect boundary conditions. Now if we perform a momentum shift by $\chi_x'=\chi_x(-1)^x$ for $e^{-iH_RT+H_IT}$ contours, then the path-integral show $O(4)$ symmetry, and again we have modes on $G^{13}_{k=\pi}(\omega,-\omega)$ and $G^{24}_{k=\pi}(\omega,-\omega)$. Note that compared to the previous case, the frequency for the modes are different. This leads to
\begin{equation}
\left<\delta G^{13}\delta G^{13}\right>= \frac{1}{\Omega^2+(k-\pi)^2}.
\end{equation}

\item The purely imaginary evolution with $J_0=J_1=0$ is also special. On the one hand, both contours are symmetric and there should be $O(4)$ symmetry. However, in this case we have $G^{12}\equiv 0$ and there should be no symmetry breaking and thus no Goldstone modes. Calculation gives
\begin{equation}
\left<\delta G^{13}\delta G^{13}\right>\sim \frac{\Theta(\omega_1\omega_2)}{k^2+|\Omega|}\ \ \ \ \ \ \ \ \ \left<\delta G^{13}\delta G^{13}\right>\sim \frac{\Theta(-\omega_1\omega_2)}{k^2+|\omega|}.
\end{equation}

\end{enumerate}

For the Brownian model, similar analysis shows for $V_0=0$ and $J_1=0$, there is a similar Goldstone mode at $G^{13}$ and $G^{24}$. However, when $V_1=0$ and $J_0=0$, the mode does not appear due to the lack of correlation in the time domain. Furthermore, different from the static model, $G^{12}$ is non-trivial even under the purely imaginary evolution. Consequently, the result is the same as the general case, as mentioned in the main text.

\end{document}